\documentclass[aps,pra,twocolumn,a4paper]{revtex4}
\usepackage{graphicx,amsmath,bm}

\def\bra#1{\mathinner{\langle{#1}|}}
\def\ket#1{\mathinner{|{#1}\rangle}}
\def\mn#1{\langle #1 \rangle}
\def\prjct#1{\mathinner{|{#1}\rangle}\!\!\mathinner{\langle{#1}|}}

\def\text#1{\textrm{#1}}

\begin{document}

\title{Detector imperfections in photon-pair source characterization}
\date{\today}
\author{P. Sekatski}
\affiliation{Group of Applied Physics, University of Geneva, CH-1211 Geneva 4, Switzerland}
\author{N. Sangouard}
\affiliation{Group of Applied Physics, University of Geneva, CH-1211 Geneva 4, Switzerland}
\author{F. Bussi\`eres}
\affiliation{Group of Applied Physics, University of Geneva, CH-1211 Geneva 4, Switzerland}
\author{C. Clausen}
\affiliation{Group of Applied Physics, University of Geneva, CH-1211 Geneva 4, Switzerland}
\author{N. Gisin}
\affiliation{Group of Applied Physics, University of Geneva, CH-1211 Geneva 4, Switzerland}
\author{H. Zbinden}
\affiliation{Group of Applied Physics, University of Geneva, CH-1211 Geneva 4, Switzerland}

\begin{abstract}
We analyze how imperfections in single-photon detectors impact the characterization of photon-pair sources. We perform exact calculations to reveal the effects of multi-pair emissions and of noisy, non-unit efficiency, non photon-number resolving detections on the Cauchy-Schwarz parameter, on the second order auto-correlation and cross-correlation functions, and on the visibilities of both Hong-Ou-Mandel and Bell-like interferences. We consider sources producing either two-mode squeezed states or states with a Poissonian photon distribution. The proposed formulas are useful in practice to determine the impacts of multi-pair emissions and dark counts in standard tests used in quantum optics.
\end{abstract}
\maketitle

\section{Introduction}
Ideally, a single-photon detector is a measurement device that perfectly discriminates the photon-number states, i.e. a box that produces $n$ clicks when the number state $|n\rangle$ comes in. In practice, however, most single-photon detectors do not resolve the photon-number, they click only once even when they interact with a bunch of photons. They are also noisy, meaning that they can click even when there is no incoming photon. Furthermore, they are often characterized by low efficiencies, especially at telecommunication wavelengths. (For a review of the current state of the art on single-photon detectors, see~\cite{Eisaman2011}.) \\

Photon sources also have their own properties, sometimes different from what one would like. To know these specificities, a battery of tests have been developed in quantum optics but most of them assume that efficient, noiseless and photon-number resolving detectors are used~\cite{Loudon2000}. \\

The goal of this paper is to make a detailed investigation of effects coming from detection imperfections on some of these tests. Specifically, we consider a photon-pair source producing a two-mode squeezed state
\begin{equation}
\label{statepsiab}
\rho_{ab}=(1-p) e^{\sqrt{p} \,a^\dag b^\dag} |00\rangle\langle 00| e^{\sqrt{p} \,ab},
\end{equation}
where $a$ and $b$ are bosonic operators corresponding to two spatial modes and $p$ is the probability of emission. This encompasses both spontaneous Raman sources in atomic ensembles as well as sources based on spontaneous four wave mixing and parametric down conversion. In particular for the latter with a pulsed pump, thermal statistics can be obtained when the photon pairs are filtered so much that the coherence time of each individual photon is larger than the pump pulse duration~\cite{Zukowski95,Bouwmeester:1997uq,PhysRevLett.80.3891,PhysRevA.67.022301,Riedmatten2004a}. (For spontaneous Raman sources with close to thermal statistics see~\cite{Kuzmich2003} and for spontaneous four wave mixing see~\cite{Li:08}.) In the continuous wave pump regime, a requirement to post-select two-mode squeezed states is to use
detectors with a time resolution shorter than the coherence time of single photons~\cite{PhysRevLett.93.070503,Halder2008}. First, we extend the definitions of the Cauchy-Schwarz parameter~\cite{Clauser1974, Kuzmich2003} in Section \ref{Cauchy_Schwarz} and of second order correlation functions~\cite{Loudon2000} in Sections \ref{Auto} and \ref{Cross} to the case where inefficient, noisy and non photon-number resolving detectors are used. We then derive their exact expressions for the state (\ref{statepsiab}) in the corresponding sections. We further consider the effect of these detection imperfections on the visibilities of a Hong-Ou-Mandel dip~\cite{HOM1987} in Section \ref{HOMdip} and of a Bell-like interference~\cite{CHSH1969} in Section \ref{Bell}. In each section, we recall the definition of the quantity under consideration as well as the experimental setup used to access it. Before the conclusion, we also consider the case of sources producing states with a Poissonian photon distribution. We believe that the proposed formulas are useful in practice to precisely estimate the effects of dark-counts and of multi-pair emissions on most of the tests that are used for the characterization of photon-pair sources. \\

\section{Model for the detector} Let us first start by describing a model associated to realistic detectors. More precisely, we are looking for the operator corresponding to a photon detector with a non-unit efficiency $\eta$, a dark count probability $p_{dc}$, which does not resolve the photon number.  A unit-efficiency and noise-free non photon-number resolving detector can perfectly discriminate the vacuum state $\ket{0}$ from all the other Fock states $\ket{n_{n\geq1}}.$ Therefore, the corresponding operator is naturally given by $\hat P_a = \mathbf{1}-\prjct{0}$ where $\mathbf{1}$ and $\prjct{0}$ stands for the identity and the projector on the vacuum, respectively (the subscript $a$ is associated to the detected mode). To include the dark counts, we apply a phase insensitive amplifier
$C_A=e^{\tanh (G) a^{\dagger}c^{\dagger}} \cosh (G)^{-a a^{\dagger}} |0_c\rangle$ on our mode. $c$ is the amplified vacuum noise and $G$ is related to the probability to get a dark count $p_{\text{dc}}$ through $p_{\text{dc}} = \tanh^2 (G).$ Therefore a noisy non photon-number resolving detector is associated with the operator
\begin{equation}
C_A^{\dagger} \hat P_a C_A = \mathbf{1}- (1-p_{\text{dc}})\prjct{0}.
\end{equation}
To account for the non-unit efficiency $\eta$, the mode $a$ is sent through a beamsplitter with transmission probability $\eta.$ The output modes are labelled $a$ and $l$ and the corresponding operator is given by $C_\eta = e^{\sqrt{(1-\eta)/\eta}\, a l^\dag} \eta^{\,\frac{1}{2} a^\dag a} \ket{0_l}.$ Note that this also accounts for a detector with an efficiency $\eta_d$ but preceded by a transmission channel with a transmission efficiency $\eta_t$ such that $\eta=\eta_d \eta_t.$ To make it short, an inefficient, noisy and non photon-number resolving detector is well modelized by the operator \footnote{The same operator can be alternatively constructed by subsequently mixing the mode with a vacuum state and a thermal state before sending it on an ideal non photon-number resolving detector.}
\begin{equation}\label{detector}
\hat D_a(\eta) = C_{\eta}^\dag C_A^\dag \hat P_a C_A C_\eta = \mathbf{1}- (1-p_{\text{dc}})(1-\eta)^{a^\dag a}.
\end{equation}
To validate our model, consider the Fock state $\ket{n}.$ The probability to get a click is $\bra{n}\hat D_a(\eta) \ket{n}= 1- (1-p_{\text{dc}})(1-\eta)^n$ which can be easily understood as one minus the probability to lose all the $n$ photons and not to get a dark count.\\
With this model in mind, we now analyze the behavior of several  tests of quantum optics that are used for the characterization of photon-pairs with realistic detectors. Note that all the detectors involved in a given setup are assumed to have the same overall efficiency but the generalization to the case where each has a different efficiency is straightforward.\\

\section{Cauchy-Schwarz parameter}\label{Cauchy_Schwarz} 
\begin{figure}
\includegraphics[width=8cm]{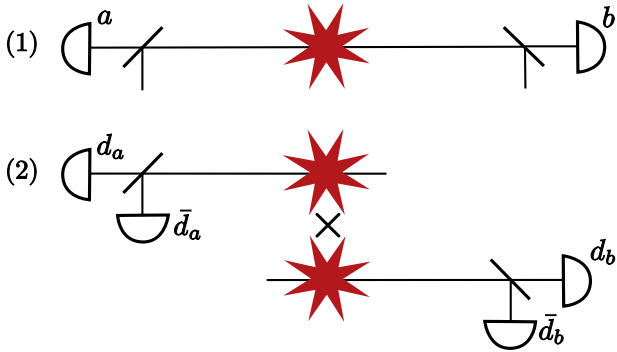}
\caption{Setup for the measurement of a Cauchy-Schwarz type of witness $\tilde{R}$ with imperfect detectors: (1) Coincidence between $a$ and $b$ with halved efficiency. (2) Product of self correlations for the two modes $a$ and $b$.}
\label{fig:R}
\end{figure}

The Cauchy-Schwarz parameter $R$~\cite{Clauser1974, Kuzmich2003} is commonly used to formally demonstrate that two fields are non-classically correlated. Let us first briefly outline the definition of classical fields. A general two-mode state $\rho_{a b}$ always admits a $P$-representation
\begin{equation}
\rho_{ab}=\int d^2\alpha\, d^2\beta P(\alpha,\beta) \prjct{\alpha, \beta},
\end{equation}
where $\alpha$ and $\beta$ are complex number and $\ket{\alpha}$, $\ket{\beta}$ are coherent states for $a$ and $b$ respectively~\cite{Mandel1995}. The state $\rho_{ab}$ is said to be classical if the quasi-probability distribution $ P(\alpha,\beta)$ is always positive. In this case, the correlation functions associated to $a$ and $b$ fulfill
\begin{equation}\label{R_factor}
R=\frac{\mn{a^\dag b^\dag b \,a}^2}{\mn{{a^{\dag 2}} {a}^2}\mn{b^{\dag^2} b^2}} \leq 1.
\end{equation}
For example, coherent state has $R=1$ while a photon-number state $|11\rangle$ has $R \rightarrow \infty.$ For a two-mode squeezed state, $R=\frac{1}{4}(1+\frac{1}{p})^2$ (see e.g.~\cite{Kuzmich2003}). The $R$ parameter is thus a witness of non-classicality but the measurement of the ratio between $\mn{a^\dag b^\dag b \,a}^2$ and the product of $\mn{a^{\dag 2} {a}^2}$ and $\mn{b^{\dag^2} b^2}$ requires noise-free photon-number resolving detectors. So, it is natural to wonder how to define this witness of non-classicality with imperfect detectors, i.e. to generalize $R$ so that it involves operators compatible with our detection model. First, let us recall that (\ref{R_factor}) is a consequence of the Cauchy-Schwarz inequality, that holds for any two measurable real-functions $f$ and $g$
\begin{equation}\label{Holder}
\left(\int |f\, g|\, d\mu\right)^2 \leq \left(\int f^2 d\mu \right)\left(\int g^2 d\mu \right).
\end{equation}
Hence, the mean value 
$$\mn{:f(a^\dag a)g(b^\dag b):}^2=\left(\text{tr} (\rho_{ab}:f(a^\dag a)g(b^\dag b):)\right)^2$$ (where by $:\,\,:$ we mean the normal ordering), which has an integral form given by
$$
\mn{:f(a^\dag a)g(b^\dag b):}^2=\left(\int  d^2 \alpha\, d^2 \beta P(\alpha,\beta) f(|\alpha|^2)g(|\beta|^2)\right)^2
$$
is bounded by
\begin{eqnarray}
\nonumber
&&\int  d^2 \alpha\, d^2 \beta P(\alpha,\beta) f^2(|\alpha|^2) \int  d^2 \alpha\, d^2 \beta P(\alpha,\beta)g(|\beta|^2)\\
\nonumber
&&=\mn{:\!f^2(a^\dag a)\!:}\mn{:\!g^2(b^\dag b)\!:}
\end{eqnarray}
for $P(\alpha,\beta) \geq 0.$
This leads to the generalized Cauchy-Schwarz inequality
\begin{equation}\label{R_general}
R_{gen}=\frac{\mn{:\!f(a^\dag a)g(b^\dag b)\!:}^2}{\mn{:\!f^2(a^\dag a)\!:}\mn{:\!g^2(b^\dag b)\!:}} \leq 1.
\end{equation}
Note that the inequality (\ref{R_factor}) is a particular case of the previous inequality where $f(x)=g(x)=x$. 
But let us focus on the case where 
\begin{equation}
\label{faa_dagpart}
f(a^\dag a) = g(a^\dag a) = 1- (1-p_{\text{dc}})(1-\eta/2)^{a^\dag a}
\end{equation}
leading to 
\begin{equation}
\label{numR}
\mn{:\!f(a^\dag a)g(b^\dag b)\!:} = \mn{\hat D_a(\eta/2) \hat D_b(\eta/2)}. 
\end{equation}
Since the normal ordered form of $f(a^\dag a)$ is given by 
\begin{equation}
f(a^\dag a)=1-(1-p_{\text{dc}}):e^{-\eta/2 a^{\dag}a}:
\end{equation}
its square is given by 
\begin{equation}
\label{denR}
f^2(a^\dag a) = 1 - 2 (1-p_{\text{dc}})(1-\eta/2)^{a^\dag a} + (1-p_{\text{dc}})^2(1-\eta)^{a^\dag a}.
\end{equation}
In order to give an explicit expression for $R$ as a function of operators $\hat D,$ it suffices to write $f^2$ as a function of $\hat D.$ To this end, we introduce the $d_a$ and $\bar d_a$ modes as the output modes of a beamsplitter on which $a$ is sent, c.f. Fig. \ref{fig:R}. They are related to $a$ by
\begin{equation}
a=\frac{1}{\sqrt{2}}(d_a+\bar d_a).
\end{equation}
The mean value $\mn{(1-p_{\text{dc} })(1-\eta/2)^{a^\dag a}}$ is alternatively given by $\mn{(1-p_{\text{dc}})(1-\eta)^{d_a^\dag d_a}}$ since the former corresponds to the detection of the $a$ mode after a $50\%$ transmission beamsplitter (this is also valid for the term with $\bar d$ only). The term $\mn {(1-p_{\text{dc}})^2(1-\eta)^{a^\dag a}}$ involves the total number of photons before the beamsplitter and is thus equivalent to $\mn {(1-p_{\text{dc}})^2(1-\eta)^{d^\dag d+\bar d^\dag \bar d}}$ by energy conservation. Hence, $\mn{f^2(a^\dag a)}=\mn{\hat D_{d_a}(\eta)\hat D_{\bar d_a}(\eta)}$ and the Cauchy-Schwarz $R$ parameter can be rewritten in an experimentally friendly way as
\begin{equation}
\label{R_friendly}
\tilde R = \frac{\mn{\hat D_a(\eta/2) \hat D_b(\eta/2)}^2}{\mn{\hat D_{d_a}(\eta)\hat D_{\bar d_a}(\eta)} \mn{\hat D(\eta)_{d_b}\hat D(\eta)_{\bar d_b}}}\leq 1.
\end{equation}
This proves that it is not necessary to have noise-free photon-number resolving detectors to access a Cauchy-Schwarz-type of witness. The corresponding setup is usual and is given in Fig. \ref{fig:R}. 
Note, however, that it is necessary to put a beamsplitter on the $a$ and $b$ paths to access the correct numerator of $\tilde R.$ For the state of interest (\ref{statepsiab}), the calculations of (\ref{numR}) and of the means value of (\ref{denR}) leads to (see \footnote{To derive the formulas  (\ref{R}), (\ref{g2}), (\ref{g2a|b}), (\ref{g2ab}), (\ref{V}) and (\ref{V_ent}), one needs the following relations
$$\text{tr} (\rho_{a}x^{a^\dag a})= (1-p)\sum_{n\geq 0}^\infty p^n x^n = \frac{1-p}{1-px}$$
$$\text{tr} (\rho_{ab} x^{a^\dag a+b^\dag b})= (1-p)\sum_{n\geq 0}^\infty p^n x^{2n} = \frac{1-p}{1-px^2}$$
$$\bra{0}e^{\frac{\sqrt{p}}{2} d^2}x^{d^\dag d}e^{\frac{\sqrt{p}}{2} d^{2\,\dag}} \ket{0}= \sum_{n\geq 0}^\infty \frac{(2n)!}{n!^2 2^{2n}}p^n x^{2n} = \frac{1}{\sqrt{1-px^2}}$$
})
\begin{equation}
\tilde R  =\left(\frac{1-2\frac{(1-p_\text{dc})(1-p)}{1-p(1-\eta/2)}+\frac{(1-p_\text{dc})^2(1-p)}{1-p(1-\eta/2)^2}}
{1-2\frac{(1-p_\text{dc})(1-p)}{1-p(1-\eta/2)}+\frac{(1-p_\text{dc})^2(1-p)}{1-p(1-\eta)}}\right)^2.
\label{R}
\end{equation}
Note that $\tilde R$ is always larger than 1 in the ideal case where $\eta=1-p_\text{dc}=1.$ In practice, we are usually interested in the first order development in $\eta$ and $p_\text{dc}$
\begin{equation}
\tilde R \approx \left( 1+(\frac{1}{2 n_a}-\frac{\eta}{4})(1-2 \frac{p_\text{dc}}{\eta n_a})\right)^2,
\end{equation}
where $n_a=\frac{p}{1-p}$ is the mean number of photons for the mode $a,$ c.f.~below.
Under the reasonable assumption $2 p_\text{dc}< \eta n_a,$ non-classical correlations can be observed as soon as $\eta n_a < 2$. 
\begin{figure}
\includegraphics[width=8cm]{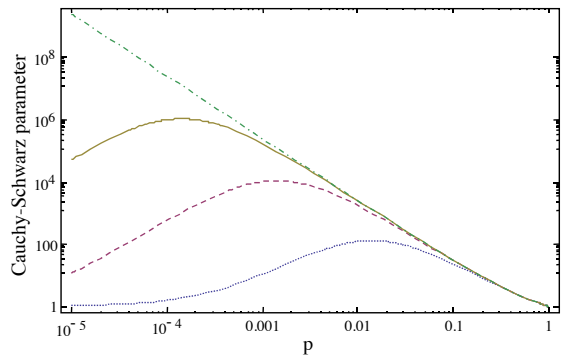}
\caption{Cauchy-Schwarz parameter $\tilde R$ (\ref{R}) for a two-mode squeezed state as a function of the emission probability $p$ for dark count probability (i) $p_{\text{dc}}=10^{-6}$ (full line) (ii) $p_{\text{dc}}=10^{-5}$ (dashed line) and (iii) $p_{\text{dc}}=10^{-4}$ (dotted line). The dashed-dotted line is the ideal $R$ for a two-mode squeezed state. The overall detection efficiency is fixed at $\eta=10^{-2}.$}
\label{fig:R_ex}
\end{figure}
Fig. \ref{fig:R_ex} shows several values of $\tilde R$ as a function of the probability of emission $p$ for fixed detection efficiency $\eta$ and for various values of the dark count probability $p_{\text dc}.$ Note that for the typical experimental values $p=0.1,$ $p_{dc}=10^{-6},$ $\eta=10^{-2},$ $\tilde R$ is about $30.$ (Note that $\eta$ is the overall detection efficiency. For other experimental values, we invite the reader to use the formula (\ref{R}).) However, in contrast to the ideal case (perfect detections) where $R=\frac{1}{4}\left(1+\frac{1}{p}\right)^2,$ there is an optimal value $p_{\text{opt}} \neq 0$ that maximizes (\ref{R}) for a given detector $\{\eta,p_{\text dc}\}.$ One finds that $p_{\text{opt}} \approx \frac{p_{\text{dc}}}{\eta}.$ In our example, this translates into $p_{\text{opt}} \approx 10^{-4}$ which gives $R_{\max}=10^{6}.$  \\

\section{Auto-correlation function}\label{Auto}  
\begin{figure}
\includegraphics[width=6cm]{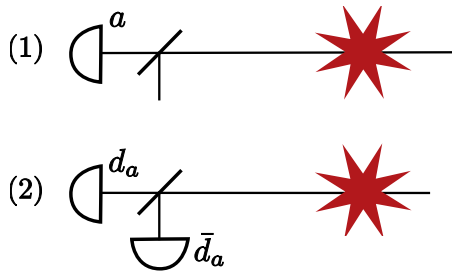}
\caption{Setup for the measurement of an auto-correlation type of function $\tilde g^{(2)}_a$ for a non-conditioned field (\ref{staterhoa}): (1) Singles associated to the mode $a$ with a halved efficiency $(\mn{D_a(\eta/2)}).$ (2) Coincidence between the two detectors $d_a$ and $\bar d_a$ $(\mn{D_d(\eta)D_{\bar d}(\eta)}).$}
\label{fig:g2}
\end{figure}

\begin{figure}
\includegraphics[width=8cm]{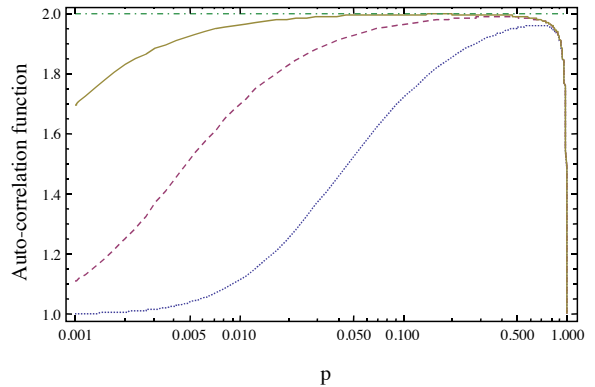}
\caption{Auto-correlation function $\tilde g^{(2)}_a$ for the non-conditioned field (\ref{staterhoa}) as a function of the emission probability $p$ for dark count probability (i) $p_{\text{dc}}=10^{-6}$ (full line) (ii) $p_{\text{dc}}=10^{-5}$ (dashed line) and (iii) $p_{\text{dc}}=10^{-4}$ (dotted line). The overall detection efficiency is fixed at $\eta=10^{-2}.$ The auto-correlation function of a thermal state is expected to be equal to 2 in the ideal case (dashed-dotted line).}
\label{fig:g2_ex}
\end{figure}

The second-order zero-time autocorrelation function $g^{(2)}$ is a witness of non-classicality for single-mode fields. Consider a single-mode field, described say by the state $\rho_a.$ It is said classical if the quasi-probability distribution $P(\alpha)$ such that
\begin{equation}
\rho_a=\int d^2\alpha P(\alpha) |\alpha\rangle \langle \alpha |
\end{equation}
is always positive. In this case, the corresponding $g^{(2)}_a$ function, defined as
\begin{equation}
\label{def_g2}
g^{(2)}_a=\frac{\langle a^{\dagger 2} a^2 \rangle}{\langle a^{\dagger} a \rangle^2},
\end{equation}
is larger (or equal) than 1. For example, coherent light has $g^{(2)}_a=1,$ thermal light has $g^{(2)}_a=2$ while a single-photon number state has $g^{(2)}_a=0$~\cite{Loudon2000}. However, how can one access such a witness with imperfect detectors? By setting $g(b^\dag b)=\mathbf 1$ in the calculation developed in the previous section, one immediately finds the following generalized inequality for classical single-mode fields 
\begin{equation}
\label{def_g2_gen}
g^{(2)}_{a, gen}=\frac{\mn{:\! f^2(a^\dag a)\!:}}{\mn{:\! f(a^\dag a) \!:}^2} \geq 1.
\end{equation}
In the specific case (\ref{faa_dagpart}), this leads to  
\begin{equation}
\tilde g^{(2)}_a=  \frac{\mn{\hat D_{d_a}(\eta) \hat D_{\bar d_a}(\eta)}}{\mn{\hat D_a(\eta/2)}^2} \geq 1.
\end{equation}
Hence, it is not necessary to have noise-free photon-number resolving detectors to access $g^{(2)}_a$-like function. When imperfect detectors are used, $\tilde g^{(2)}_a$ has to be measured by sending the state $\rho_a$ onto a beamsplitter and by taking the ratio between the number of two-fold coincidences and the number of singles squared after the beamsplitter, as shown in Fig.~\ref{fig:g2}. \\
Consider first one of the two individual modes of the two-mode squeezed state
\begin{equation}
\label{staterhoa}
\rho_{a}=\text{tr}_b \, \rho_{ab}= (1-p)  \sum_{n=0}^{+\infty} p^{n} |n\rangle\langle n|=\text{tr}_a \, \rho_{ab}=\rho_b.
\end{equation}
Using the results presented in the previous section, one finds for the specific state (\ref{staterhoa})
\begin{equation}\label{g2}
\tilde g^{(2)}_a=\frac{1-2\frac{(1-p_\text{dc})(1-p)}{1-p(1-\eta/2)}+\frac{(1-p_\text{dc})^2(1-p)}{1-p(1-\eta)}}
{\left(1-\frac{(1-p_\text{dc})(1-p)}{1-p(1-\eta/2)}\right)^2}.
\end{equation} 
Note that the mean number of photons in the state $\rho_a$ is $n_a=\frac{p}{1-p}$ (which is well approximated by $p$, if the later is small). Developing $\tilde g^{(2)}$ in Taylor series for $p_\text{dc}$ leads to the simple expression  
\begin{equation}
\tilde g^{(2)}_a= \left(2-\frac{\eta n_a}{1+\eta n_a}\right)\left(1- 2 \frac{p_\text{dc}}{\eta n_a}+O[p_\text{dc}]^2\right).
\end{equation}
For the typical experimental values $p=0.1,$ $p_{dc}=10^{-6},$ $\eta=10^{-2},$ the auto-correlation function $\tilde g^{(2)}$ is about $1.996,$ i.e. very close to the expected value with ideal detections. Fig. \ref{fig:g2_ex} shows that the auto-correlation function is about $2$ as long as the detector noise is small $p_{\text{dc}} \ll 10^{-2}$ and negligible with respect to the signal $(\approx p\eta).$ When the noise dominates, the auto-correaltion function is reduced to $1$ as expected from uncorrelated noises.\\
It is also interesting to evaluate the auto-correlation function of the conditioned field $\rho_{a|b}$, i.e. the field describing the mode $a$ when the twin mode $b$ is detected~\cite{Grangier1986}.
\begin{eqnarray}
\label{rhoa|b}
&\rho_{a|b}&=\frac{1}{\text{tr}_a \rho_{a|b}}\text{tr}_b \left(\hat D_b(\eta) \rho_{ab}\right) \\
\nonumber
&&= \frac{(1-p)}{\text{tr}_a \rho_{a|b}} \sum_{n=0}^{+\infty} p^n (1-(1-p_{\text{dc}})(1-\eta)^n)|n\rangle\langle n|,
\end{eqnarray}
with $\text{tr}_a \rho_{a|b}=1-\frac{(1-p_{\text{dc}})(1-p)}{1-p(1-\eta)}$ and its auto-correlation function is given by
\begin{equation}
\label{g2a|b}
\tilde g^{(2)}_{a|b}=\frac{1-2(1-p_{\text{dc}})\zeta(1-\eta/2)+(1-p_{\text{dc}})^2\zeta(1-\eta)}{\left(1-(1-p_{\text{dc}})\zeta(1-\eta/2)\right)^2}
\end{equation}
where 
\begin{equation}
\zeta(x)=\frac{1}{\text{tr}_a \rho_{a|b}}\left(\frac{1-p}{1-px}-\frac{(1-p_{\text{dc}})(1-p)}{1-p(1-\eta)x}\right).
\end{equation}
\begin{figure}
\includegraphics[width=8cm]{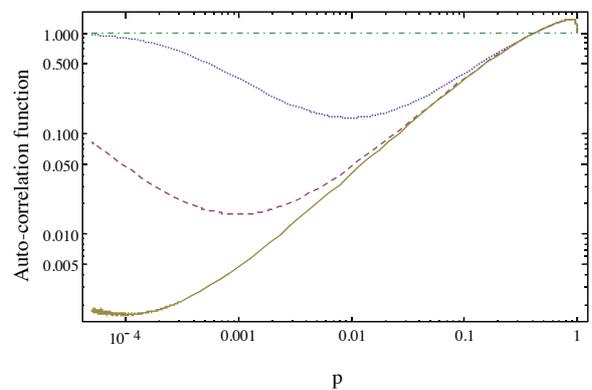}
\caption{Auto-correlation function $\tilde g^{(2)}_{a|b}$ for the conditioned state (\ref{rhoa|b}) as a function of the emission probability $p$ for dark count probability (i) $p_{\text{dc}}=10^{-6}$ (full line) (ii) $p_{\text{dc}}=10^{-5}$ (dashed line) and (iii) $p_{\text{dc}}=10^{-4}$ (dotted line). The dashed-dotted line is the non-classical threshold. The overall detection efficiency is fixed at $\eta=10^{-2}.$}
\label{fig:g2a|b_ex}
\end{figure}
Fig. \ref{fig:g2a|b_ex} shows that $\tilde g^{(2)}_{a|b}$ reveals the non-classicality of the conditioned field provided that $p$ is small enough and that the signal $p\eta$ dominates over the noise $p_{\text{dc}}.$ It also appears that for fixed $\{\eta,p_{\text{dc}}\},$ $\tilde g^{(2)}_{a|b}$ is minimized for $p_{\text{opt}}=\frac{p_{\text{dc}}}{\eta}.$ For example, for $p_{\text{dc}}=10^{-6}$ and $\eta=10^{-2},$ one gets $\tilde g^{(2)}_{a|b, \, \min}=10^{-3}$ for $p_{\text{opt}}=10^{-4}.$ \\

\section{Cross-correlation function}\label{Cross}
\begin{figure}
\includegraphics[width=8cm]{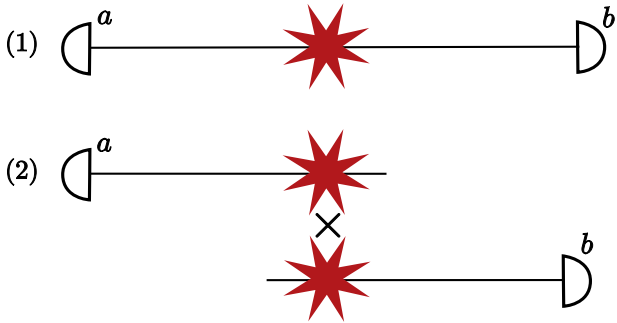}
\caption{Setup for the measurement of the cross-correlation function $\tilde g^{(2)}_{ab}$: (1) Two-fold coincidence measurement. (2) Measurements of the individual singles.}
\label{fig:crosscorr}
\end{figure}

The second order cross-correlation function, defined ideally as
\begin{equation}
\label{def_gab}
g_{ab}^{(2)}=\frac{\langle a^{\dagger} b^{\dagger} b a\rangle}{\langle a^{\dagger} a \rangle \langle b^{\dagger} b \rangle},
\end{equation}
is exploited to get information about the photon-number distribution of a two-mode field. From its ideal definition, the cross-correlation function can in principle be directly deduced  from $R$ and $g^{(2)}$ using 
\begin{equation}
g_{ab}^{(2)}=\sqrt{R \-\ g^{(2)}_a g^{(2)}_b}.
\end{equation}
However, when one uses non photon-number resolving detectors, we have seen that the measurement of mean values of the form $\langle a^{\dag 2} a^2 \rangle$ requires a beamsplitter and two detectors. The beamsplitter cannot be removed for the measurements of both $\langle a^{\dag} b^{\dag} b a \rangle$ or $\langle a^{\dag} a \rangle$ when one wants to determine the Cauchy-Schwarz parameter and the auto-correlation function, respectively. But the beamsplitter is not necessary for the measurement of $g_{ab}^{(2)}$ since it does not involve terms of the form  $\langle a^{\dag 2} a^2 \rangle.$ $g_{ab}^{(2)}$ is merely measured by placing one detector on each of the paths $a$ and $b$ from the ratio between the two-fold coincidences and the product of singles, c.f. Fig. \ref{fig:crosscorr}. 
Therefore
\begin{equation}
\tilde g^{(2)}_{ab}=\frac{\mn{D_a(\eta)D_b(\eta)}}{\mn{D_a(\eta)}\mn{D_b(\eta)}},
\end{equation}
which leads to
\begin{equation}
\label{g2ab}
\tilde g^{(2)}_{ab}=\frac{1-2\frac{(1-p_{\text{dc}})(1-p)}{1-p(1-\eta)}+\frac{(1-p_{\text{dc}})^2(1-p)}{1-p(1-\eta)^2}}{\left(1-\frac{(1-p_{\text{dc}})(1-p)}{1-p(1-\eta)}\right)^2}.
\end{equation}
for the state (\ref{statepsiab}).
Note that for $p_{\text{dc}}=0,$ the previous expression takes a very simple form
\begin{equation}
\tilde g^{(2)}_{ab}=1+\frac{1}{p}\left(\frac{1-p}{1-p(1-\eta)^2}\right).
\end{equation}
\begin{figure}
\includegraphics[width=8cm]{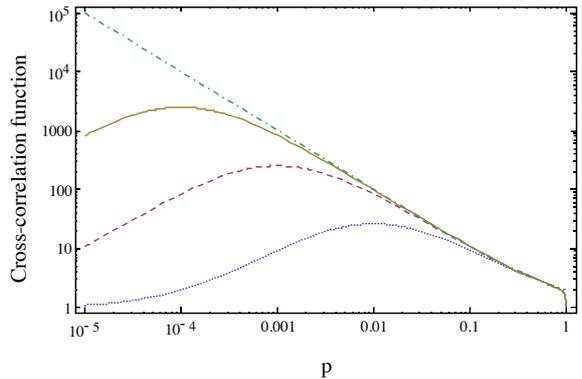}
\caption{Cross-correlation function $\tilde g^{(2)}_{ab}$ for a two-mode squeezed state as a function of the emission probability $p$ for dark count probability (i) $p_{\text{dc}}=10^{-6}$ (full line) (ii) $p_{\text{dc}}=10^{-5}$ (dashed line) and (iii) $p_{\text{dc}}=10^{-4}$ (dotted line). The overall detection efficiency is fixed at $\eta=10^{-2}.$ The dashed-dotted line gives the cross-correlation function that would be ideally obtained for a two-mode squeezed state $g^{(2)}_{ab}=1+\frac{1}{p}$ (see e.g.~\cite{RMP2011}).}
\label{fig:g2ab_ex}
\end{figure}
Further note that for $p=0.1,$ $p_{dc}=10^{-6},$ $\eta=10^{-2},$ one finds $\tilde g^{(2)}_{ab} \approx 11.$ However, Fig. \ref{fig:g2ab_ex} shows that there is an optimal value $p_{\text{opt}} \approx \frac{p_{\text{dc}}}{\eta}$ for given $\{\eta,p_{\text{dc}}\}.$ For the previous example, $p_{\text{opt}} \approx 10^{-4}$ leading to $\tilde g^{(2)}_{ab, \, \max} \approx 2500.$\\

\section{Hong-Ou-Mandel interference}\label{HOMdip} 
\begin{figure}
\includegraphics[width=8cm]{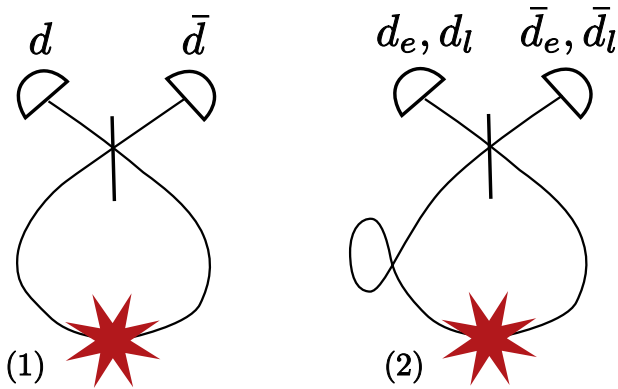}
\caption{Measurement of the Hong-Ou-Mandel interference: (1) The optical path are the same, the photons bunch. (2) One path is delayed, the two photons leads to independent modes.}
\label{fig:Dip}
\end{figure}

When two indistinguishable photons coming from the same source (or from two different sources but in pure states, see the exhaustive list of the corresponding references in~\cite{PhysRevA.81.021801}) enter a 50:50 beamsplitter, one in each input port, they coalesce. Therefore, when two photodetectors monitor the output of the beamsplitter as a function of the delay between the two input photons, the coincidence rate of the detectors drop to zero when the two photons arrived at the same time at the beamsplitter. This is known as the Hong-Ou-Mandel (HOM) interference~\cite{HOM1987}. The visibility of this interference can be used to determine the indistinguishability in all the degrees of freedom of two photons produced by the same source. But how do detection imperfections modify the visibility of a Hong-Ou-Mandel interference? We consider the experiment drawn in Fig.~\ref{fig:Dip} and we start with the state ({\ref{statepsiab}}).
When the modes $a$ and $b$ follow optical paths with the same length, they arrive at the same time at the beamsplitter and give two modes $d$ and $\bar d$ defined as
\begin{eqnarray}
\nonumber
&& a=\frac{1}{\sqrt{2}}(d+\bar d), \\
\nonumber
&& b=\frac{1}{\sqrt{2}}(d- \bar d),
\end{eqnarray}
after the beamsplitter. The state after the beamsplitter is
$\ket{\psi_{dip}} = \sqrt{1-p}\, e^{\sqrt{p}/2 (d^{\dag 2}- \bar d^{\dag 2})}\ket{00}$.
Hence, the probability of twofold coincidences is given by
$\bra{\psi_{dip}} \hat D_d(\eta) \hat D_{\bar d}(\eta) \ket{\psi_{dip}}.$
In the scenario where $a$ follows a longer optical path than $b,$ they do not arrive simultaneously at the beamsplitter and lead to four output modes
\begin{eqnarray}
\nonumber
&& a=\frac{1}{\sqrt{2}}(d_{\ell}+\bar d_{\ell}), \\
\nonumber
&& b =\frac{1}{\sqrt{2}}(d_{e}-\bar d_{e}),
\end{eqnarray}
after the beamsplitter, yielding $\ket{\psi_{out}}= \sqrt{1-p}\, e^{\sqrt{p}/2 (d^\dag_{\ell}+\bar d^\dag_{\ell})(d^\dag_{e}-\bar d^\dag_{e})}\ket{0000}$. The subscript $e$ and $\ell$ stand for early and late. In this case, each detector accounts for two temporal modes and the corresponding operator has to be modified in such a way that an overall detection event corresponds to ``at least a single click for one of the two modes $d_e$ or $d_l.$'' More formally, the probability for a coincidence is calculated from $\bra{\psi_{out}} \hat D_{d_e,d_\ell}(\eta) \hat D_{\bar d_e,\bar d_\ell}(\eta)\ket{\psi_{out}}$ where
\begin{equation}
\hat D_{d_e,d_\ell}(\eta)= \mathbf{1}- (1-p_{\text{dc}}) (1-\eta)^{d_e^\dag d_e + d_\ell ^\dag d_\ell}.
\end{equation}
Note that in this case, the noise term is not squared since the dark count probability is given for the detection gate, which is unchanged. As usual, the visibility is given by the ratio between the depth of the dip and the maximal coincidence rate
\begin{equation}
\tilde V_{\text{HOM}} = \frac{\bra{\psi_{out}} \hat D_{d_e,d_\ell} \hat D_{\bar d_e,\bar d_\ell}\ket{\psi_{out}}-\bra{\psi_{dip}} \hat D_d \hat D_{\bar d} \ket{\psi_{dip}}}{\bra{\psi_{out}} \hat D_{d_e,d_\ell} \hat D_{\bar d_e,\bar d_\ell}\ket{\psi_{out}}}.
\end{equation}
After some algebra, one finds
\begin{equation}
\label{V}
\tilde V_{\text{HOM}} = \frac{2(1-p_\text{dc})\left( \sqrt{\frac{1-p}{1-p(1-\eta)^2}}-\frac{1-p}{1-p(1-\eta/2)^2} \right)}
{1-2\frac{(1-p_\text{dc})(1-p)}{1-p(1-\eta/2)^2}+\frac{(1-p_\text{dc})^2(1-p)}{1-p(1-\eta)^2}}.
\end{equation}
Disregarding the dark counts ($p_\text{dc}=0$) and developing the last expression in first order with respect to $\eta,$ one has
\begin{equation}
\tilde V_{\text{HOM}} \approx \frac{1+p}{1+3p} +\frac{2p}{(1+3p)^2}\eta.
\end{equation}
\begin{figure}
\includegraphics[width=8cm]{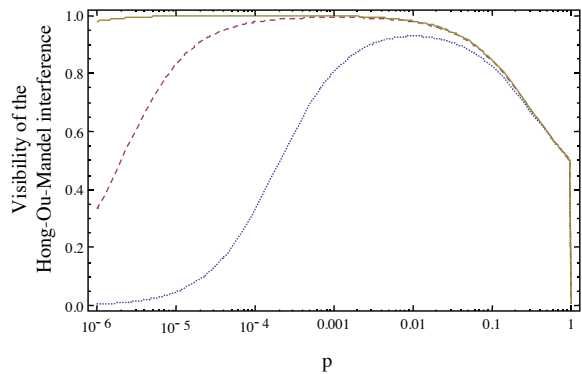}
\caption{Visibility $\tilde V_{\text{HOM}}$ of the Hong-Ou-Mandel interference for a two-mode squeezed state as a function of the emission probability $p$ for dark count probability (i) $p_{\text{dc}}=10^{-6}$ (full line) (ii) $p_{\text{dc}}=10^{-5}$ (dashed line) and (iii) $p_{\text{dc}}=10^{-4}$ (dotted line). The overall detection efficiency is fixed at $\eta=10^{-2}.$}
\label{fig:Dipex}
\end{figure}
For the typical values $p=0.1,$ $p_{dc}=10^{-6},$ $\eta=10^{-2},$ $\tilde V_{\text{HOM}}$ is about $0.85$. Fig. \ref{fig:Dipex} shows the behavior of $\tilde V_{\text{HOM}}$ as a function of $p$ for various detection noise. The maximal value is obtained when $p\ll1$ and $p\eta \gg p_{\text{dc}}.$ In our example, $\tilde V_{\text{HOM,} \, \max} \approx 0.98$ for $p \approx 10^{-2}.$ \\

\section{Bell interference}\label{Bell}
\begin{figure}
\includegraphics[width=8cm]{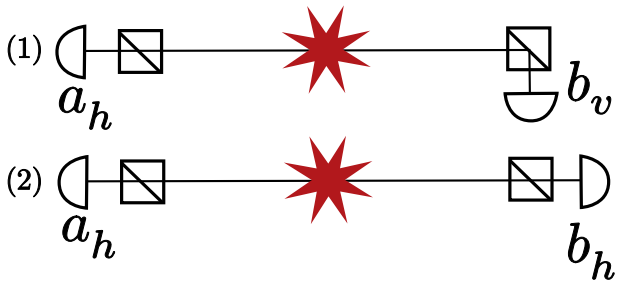}
\caption{Measurement of the Bell interference: (1) Alice and Bob measure in maximally correlated bases. (2) Alice and Bob measure in minimally correlated bases.}
\label{fig:EPR}
\end{figure}

Bell inequalities have initially been proposed to test quantum non-locality. However, all non-local states are entangled. Therefore, a Bell test can also be used as a witness of entanglement. Let us recall the principle of a Bell interference. Two distant persons, Alice, located at the location A, Bob at B, share a quantum state $\rho_{ab}$. Alice chooses one setting among $\{\sigma_x, \sigma_z\}$ while Bob rotates his basis measurement in the $(xz)$ plan. They then record the number of coincidences as a function of the angle between Alice's and Bob's settings for each of Alice's measurements. The only state that can produce a visibility of 100\% is a two-partite maximally-entangled state. Under the assumption that the state is a mixture between a maximally entangled state and white noise, one can conclude about the presence of entanglement with the Clauser-Horne-Shimony-Holt inequality~\cite{CHSH1969} if the average visibility is larger than $1/\sqrt{2}.$ But what is the visibility resulting from a Bell interference from a source producing say, polarization-entangled pairs,
\begin{equation}
\label{ent}
\ket{\psi_{ab}^-}= (1-p)  e^{\sqrt{p} \,(a_h^\dag b_v^\dag-a_v^\dag b_h^\dag)} |00\rangle
\end{equation}
and characterized with imperfect detectors? The setup is given in Fig. \ref{fig:EPR}. Consider the case where Alice chooses $\sigma_z$ and focuses on the clicks she gets along the horizontal polarization. The maximal number of coincidences between Alice's and Bob's detector is obtained from
\begin{equation}
\nonumber
N^c_{hv} = \bra{\psi_{ab}^-} \hat D_{a_h} \hat D_{b_v}\ket{\psi_{ab}^-}
\end{equation}
while the minimal number of coincidences is given by
\begin{equation}
\nonumber
N^c_{hh} = \bra{\psi_{ab}^-} \hat D_{a_h} \hat D_{b_h}\ket{\psi_{ab}^-}
\end{equation}
The visibility of the interference is given by
\begin{eqnarray}
\label{V_ent}
\tilde V_{\text{ent}}=\frac{N^c_{hv}-N^c_{hh}}{N^c_{hv}+N^c_{hh}}=\qquad\qquad\qquad\qquad\\ \nonumber
\frac{\frac{(1-p_\text{dc})^2(1-p)}{1-p(1-\eta)^2}-\frac{(1-p_\text{dc})^2(1-p)^2}{(1-p(1-\eta))^2}}
{2-4 \frac{(1-p_\text{dc})(1-p)}{1-p(1-\eta)}+\frac{(1-p_\text{dc})^2(1-p)}{1-p(1-\eta)^2}+\frac{(1-p_\text{dc})^2(1-p)^2}{(1-p(1-\eta))^2}}.
\end{eqnarray}
Again, if we disregard the dark counts $(p_{\text{dc}}=0),$ this expression reduces to
\begin{equation}
\tilde V_{\text{ent}} = \frac{1-p}{1+p-2p^2(1-\eta)^2}.
\end{equation}
\begin{figure}
\includegraphics[width=8cm]{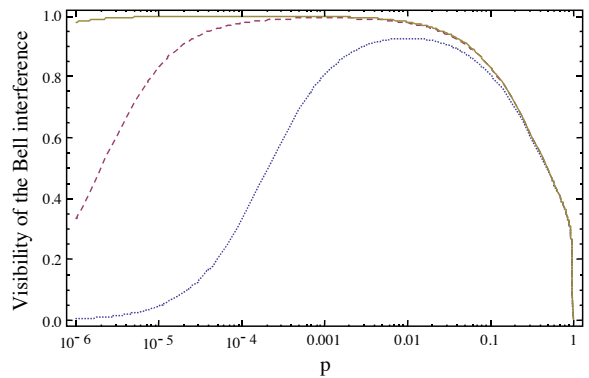}
\caption{Visibility $\tilde V_{\text{ent}}$ of Bell-type interference for the state (\ref{ent}) as a function of the emission probability $p$ for dark count probability (i) $p_{\text{dc}}=10^{-6}$ (full line) (ii) $p_{\text{dc}}=10^{-5}$ (dashed line) and (iii) $p_{\text{dc}}=10^{-4}$ (dotted line). The overall detection efficiency is fixed at $\eta=10^{-2}.$}
\label{fig:Bellex}
\end{figure}
For the typical values $p=0.1,$ $p_{dc}=10^{-6},$ $\eta=10^{-2},$ $\tilde V_{\text{ent}}$ is about $0.83$. Fig. \ref{fig:Bellex} shows the behavior of $\tilde V_{\text{ent}}$ as a function of $p.$ This behavior is similar to the one of $\tilde V_{\text{HOM}}.$ In particular, the maximal value is also obtained when $p\ll1$ and $p\eta \gg p_{\text{dc}}.$ In our example, $\tilde V_{\text{ent,} \, \max} \approx 0.98$ for $p_{\text{opt}} \approx 10^{-2}.$ \\

\section{Multimode case}
So far, we have considered the case where the source produces two-mode squeezed states. However, when the generated photons are not strongly filtered, the duration time of the pump field can be longer than the coherence time of individual photons (for spontaneous parametric down conversion see e.g.~\cite{PhysRevLett.25.84,tanzilli:26,PhysRevLett.86.5620}, for spontaneous four wave mixing see e.g.~\cite{Fiorentino2002a,Rarity:05,Fan:05,Takesue:05,Dyer:08,Slater:10}). In this case, the created pairs are well described by 
\begin{equation}
\label{state_th}
\rho_{ab}^N=(1-\bar p)^N e^{\sqrt{\bar p} \sum_{n=1}^{N} a_n^{\dag}b_n^{\dag}} |0\rangle\langle0|  e^{\sqrt{\bar p} \sum_{n=1}^{N} a_n b_n} 
\end{equation}
where $N$ is the number of modes corresponding to the ratio of coherence times between the pump and the photons. We introduced the emission probability $\bar p$ such that the average number of photons is given by $N \frac{\bar p}{1- \bar p}.$ If one wants to compare the results for the thermal and Poissonian photon distributions, the average number of photons has to be identical, that is $\bar p = \frac{p}{N-p(N-1)}.$ For a detector corresponding to 
\begin{equation}
\bar D_{\sum_n a_n}(\eta)= \mathbf{1}-(1-p_{dc})(1-\eta)^{\sum_n a_n^{\dag}a_n}
\end{equation}  
one finds \footnote{The extension of our calculations to states of the form (\ref{state_th}) is easy if ones notes that 
$$
\text{tr} (\rho_{ab}^N x^{\sum_{n=0}^N a_n^\dag a_n})= \Pi_{n=0}^N \text{tr} \rho_{a_nb_n} x^{a_n^{\dag}a_n}= \left(\frac{1-\bar p}{1-\bar px}\right)^N
$$
where $\rho_{a_nb_n} $ is given from the Eq. \ref{statepsiab}) by replacing $a, \-\ b$ by $a_n, \-\ b_n$ and $p$ by $\bar p$. Similarly
$$
\text{tr} (\rho_{ab}^N x^{\sum_{n=0}^{N} a_n^\dag a_n+b_n^\dag b_N}) = \left(\frac{1-\bar p}{1-\bar px^2}\right)^N. 
$$}
\begin{eqnarray}
\nonumber
&&\tilde R  =\left(\frac{1-2\frac{(1-p_\text{dc})(1-\bar p)^N}{\left(1-\bar p(1-\eta/2)\right)^N}+\frac{(1-p_\text{dc})^2(1-\bar p)^N}{\left(1-\bar p(1-\eta/2)^2\right)^N}}
{1-2\frac{(1-p_\text{dc})(1-\bar p)^N}{\left(1-\bar p(1-\eta/2)\right)^N}+\frac{(1-p_\text{dc})^2(1-\bar p)^N}{\left(1-\bar p(1-\eta)\right)^N}}\right)^2,\\
\nonumber
&& \tilde g^{(2)}_{\sum_n a_n}=\frac{1-2\frac{(1-p_\text{dc})(1-\bar p)^N}{\left(1-\bar p(1-\eta/2)\right)^N}+\frac{(1-p_\text{dc})^2(1-\bar p)^N}{\left(1-\bar p(1-\eta)\right)^N}}
{\left(1-\frac{(1-p_\text{dc})(1- \bar p)^N}{\left(1-\bar p(1-\eta/2)\right)^N}\right)^2}.
\end{eqnarray}
Furthermore, $\tilde g_{\sum_n a_n |\sum_n b_n}^{(2)}$ is still given by (\ref{g2a|b}) but with
 \begin{equation}
 \nonumber
\zeta(x)=\frac{\left(\frac{1-\bar p}{1-\bar px}\right)^N-\frac{(1-p_{\text{dc}})(1-\bar p)^N}{\left(1-\bar p(1-\eta)x\right)^N}}{1-\frac{(1-p_{\text{dc}})(1-\bar p)^N}{\left(1-\bar p(1-\eta)\right)^N}}.
 \end{equation}
 One also finds
$$
\tilde g^{(2)}_{\sum_n a_nb_n}=\frac{1-2\frac{(1-p_{\text{dc}})(1-\bar p)^N}{\left(1-\bar p(1-\eta)\right)^N}+\frac{(1-p_{\text{dc}})^2(1-\bar p)^N}{\left(1-\bar p(1-\eta)^2\right)^N}}{\left(1-\frac{(1-p_{\text{dc}})(1-\bar p)^N}{\left(1-\bar p(1-\eta)\right)^N}\right)^2},
$$
\newpage
$$
\tilde V_{\text{HOM}} = \frac{2(1-p_\text{dc})\left(\left(\frac{1-\bar p}{1-\bar p(1-\eta)^2}\right)^{N/2}-\left(\frac{1-\bar p}{1-\bar p(1-\eta/2)^2}\right)^N \right)}
{1-2\frac{(1-p_\text{dc})(1- \bar p)^N}{\left(1-\bar p(1-\eta/2)^2\right)^N}+\frac{(1-p_\text{dc})^2(1-\bar p)^N}{\left(1-p(1-\eta)^2\right)^N}},
$$
\begin{eqnarray}
\nonumber
&&\tilde V_{\text{ent}}=\left(\frac{(1-p_\text{dc})^2(1-\bar p)^N}{\left(1-\bar p(1-\eta)^2\right)^N}-\frac{(1-p_\text{dc})^2(1-\bar p)^{2N}}{(1-\bar p(1-\eta))^{2N}}\right) \times\\
\nonumber
&& \frac{1}{2-4 \frac{(1-p_\text{dc})(1-\bar p)^N}{\left(1-\bar p(1-\eta)\right)^N}+\frac{(1-p_\text{dc})^2(1-\bar p)^N}{\left(1-p(1-\eta)^2\right)^N}+\frac{(1-p_\text{dc})^2(1-\bar p)^{2N}}{\left(1-p(1-\eta)\right)^{2N}}}.
 \end{eqnarray}

\section{Conclusion}
We have analyzed how imperfections in single-photon detectors impact the characterization of photon-pair sources.The formulas (\ref{R}), (\ref{g2}), (\ref{g2a|b}), (\ref{g2ab}), (\ref{V}) and (\ref{V_ent}) give the Cauchy-Schwarz parameter, the second-order auto-correlation function for both non-conditioned and conditioned states, the cross-correlation function, the visibility of the Hong-Ou-Mandel dip and the visibility of Bell-like interference respectively, when inefficient, noisy and non photon-number detections are performed on two-modes squeezed states. They are exact in the sense that they are valid even for large probabilities of emissions. The previous section extend all these formulas in the case of a multimode emission. We believe that all these formulas will be useful for the experimentalists who wants to know the effect of detector noise and of multi-pair emissions on the tests they used to characterize their pair sources.

We thank A. Martin and S. Tanzilli who provide us the motivations to write this paper. We also thank them, as well as M. Afzelius and C.I. Osorio for interesting discussions. We gratefully acknowledge support by the EU projects Qscale and Qessence and from the Swiss NCCR QSIT.

\bibliography{bib}

\end{document}